\documentclass{webofc}

\usepackage[varg]{txfonts}   
\usepackage{hyperref}
\usepackage{url}
\usepackage[nameinlink,capitalise,noabbrev]{cleveref} 
\usepackage{subcaption} 

\newcommand{\muEQCD}{\mu_{\mathrm{EQCD}}}

\newcommand{\nf}{n_f}
\hypersetup{colorlinks=true,citecolor=blue,urlcolor=blue,linkcolor=blue}
\begin{document}
\title{Transport Coefficients from pQCD to the Hadron Resonance Gas at finite BSQ densities}
%
%

\author{\firstname{Isabella} \lastname{Danhoni}\inst{1}\fnsep\thanks{\email{idanhoni@illinois.edu}}}
\institute{University of Illinois Urbana–Champaign}

\abstract{We calculate the shear viscosity, $\eta$, in two limits: perturbative QCD and an excluded-volume hadron resonance gas (HRG), at finite BSQ densities. Using an interpolation framework, we connect these regimes. In addition, we present results for (almost) next-to-leading order weak-coupling shear viscosity for QCD at finite $\mu_B$, and discuss the convergence of the perturbative series
}

\maketitle

\vspace{-0.9cm}
\section{Introduction}
\label{intro}
\vspace{-0.2cm}
Heavy-ion collisions explore the deconfined state of matter composed of quarks and gluons, known as the quark–gluon plasma (QGP). Transport coefficients, such as the shear viscosity $\eta$, are fundamental properties of the QGP and encode key information on the deconfined phase. As a function of temperature, the specific shear viscosity $\eta/s$ is generally believed to have a minimum near the transition between the confined and deconfined phases, and increases at high temperatures, in the QGP phase, and at low temperatures in the hadron resonance gas (HRG) phase. In this work, we explore shear viscosity in the finite-density region of the QCD phase diagram and provide a description of the kinematic ratio $\eta T/w\!\left(T,\mu_B,\mu_S,\mu_Q\right)$ (with $w=\varepsilon+p$ the enthalpy density), which reduces to $\eta/s$ when the chemical potentials vanish. For that, we employ perturbative QCD at high temperatures and a hadron resonance gas at low temperatures, and connect the two regimes with an interpolation function.
\vspace{-0.3cm}
\section{Shear Viscosity across the QCD phase diagrams}
\label{sec-2}
\vspace{-0.2cm}
At low $T$, we employ the HRG model with repulsive interactions implemented via an excluded volume (EV-HRG) approach. We take a common eigenvolume for all species, i.e., $v=v_i$ for all $i$. We begin by defining the effective (dimensionless) chemical potential for species $i$ is
\begin{equation}
    \tilde{\mu}_i = B_i \frac{\mu_B}{T} + S_i \frac{\mu_S}{T} + Q_i \frac{\mu_Q}{T}.
    \label{eqn:effchem}
\end{equation}
For an excluded-volume description, the pressure satisfies the self-consistent relation~\cite{Rischke:1991ke,Yen:1997rv,Gorenstein:1999ce,Vovchenko:2016ebv}
\begin{align}
\nonumber
    \frac{p^{\mathrm{ex}}(T,\mu_B,\mu_S,\mu_Q)}{T}
    &= n^{\mathrm{id}}(T,\mu_B,\mu_S,\mu_Q)\,
       \exp\!\left(-\,\frac{v\, p^{\mathrm{ex}}(T,\mu_B,\mu_S,\mu_Q)}{T}\right),
\end{align}
where $n^{\mathrm{id}}\equiv \sum_i n_i^{\mathrm{id}}$. The remaining thermodynamic quantities follow from standard thermodynamic relations. These calculations are performed with \textsc{Thermal-FIST}~\cite{Vovchenko:2019pjl} using the PDG2021+ list~\cite{SanMartin:2023zhv}.
At finite chemical potentials, the HRG shear viscosity takes the form (see~\cite{Danhoni:2024kgi} for details)

\begin{equation}
    \eta^\mathrm{HRG} = \frac{5}{64\sqrt{8}}\frac{1}{r^2}\frac{1}{n_\mathrm{tot}^\mathrm{id}}\sum_i n_i^\mathrm{id}\frac{\int_0^\infty  k^3 \exp{\left(\frac{-\sqrt{k^2 + m_i^2}}{T}\right)} dk}{\int_0^\infty  k^2 \exp{\left(\frac{-\sqrt{k^2 + m_i^2}}{T}\right)} dk},
    \label{shear_hrg}
\end{equation}
where $n_{\mathrm{tot}}=\sum_i n_i$, and the kinematic ratio is
\begin{equation}
    \frac{\eta^{\mathrm{HRG}}\, T}{\varepsilon^{\mathrm{ex}} + p^{\mathrm{ex}}}
    = \frac{\eta^{\mathrm{HRG}}\, T}{w^{\mathrm{ex}}}\,,
\end{equation}
with $w^{\mathrm{ex}}=\varepsilon^{\mathrm{ex}}+p^{\mathrm{ex}}$ the enthalpy density.

For the high-$T$ regime, we use the framework of Refs.~\cite{Arnold:2000dr,Arnold:2003zc}, later extended to finite $\mu_B$ in Ref.~\cite{Danhoni:2022xmt}. We consider the system close to local equilibrium, so the non-equilibrium distribution for species $a$ (gluon, quark, or antiquark) is written as
\begin{equation}
\label{f1}
    f^a(\vec k,\vec x) = f_0^a(\vec k,\vec x) + f_0^a\!\left(1 \pm f_0^a\right)\, f_1^a(\vec k,\vec x)\,.
\end{equation} 
where the local-equilibrium distribution functions are $f_0^{q/\bar q}=[\exp(p/T-\tilde\mu_i)+1]^{-1}$ and $f_0^g=[\exp(p/T)-1]^{-1}$.
A Boltzmann-type equation governs the dynamics of $f^a$. Neglecting time derivatives and external forces, we have
\begin{equation}
    \vec v_k \cdot \frac{\partial}{\partial \vec x}\, f^a(\vec k,\vec x,t) = -\,\mathcal{C}^a[f]\,,
    \label{boltz}
\end{equation}
with the collision operator given by
\begin{align}
    \nonumber
    \mathcal{C}^a[f](\vec p) &= \frac{1}{2} \sum_{bcd}
    \int_{\vec k,\vec p',\vec k'} 
    \frac{|\mathcal{M}_{abcd}(P,K,P',K')|^2}{2p^0\, 2k^0\, 2p'{}^0\, 2k'{}^0}\,
    (2\pi)^4 \delta^{(4)}(P+K-P'-K') \\
    & \times \Big\{ f^a(\vec p)\,f^b(\vec k)\,[1\!\pm\! f^c(\vec p')]\,[1\!\pm\! f^d(\vec k')]
      - f^c(\vec p')\,f^d(\vec k')\,[1\!\pm\! f^a(\vec p)]\,[1\!\pm\! f^b(\vec k)] \Big\},
    \label{collop22}
\end{align}
where $\mathcal{M}$ denotes the QCD scattering matrix element; see Ref.~\cite{Danhoni:2024kgi} for details.
\begin{figure}[htbp]
  \centering
  \begin{minipage}{0.4\columnwidth}
    \centering
    \includegraphics[width=\linewidth]{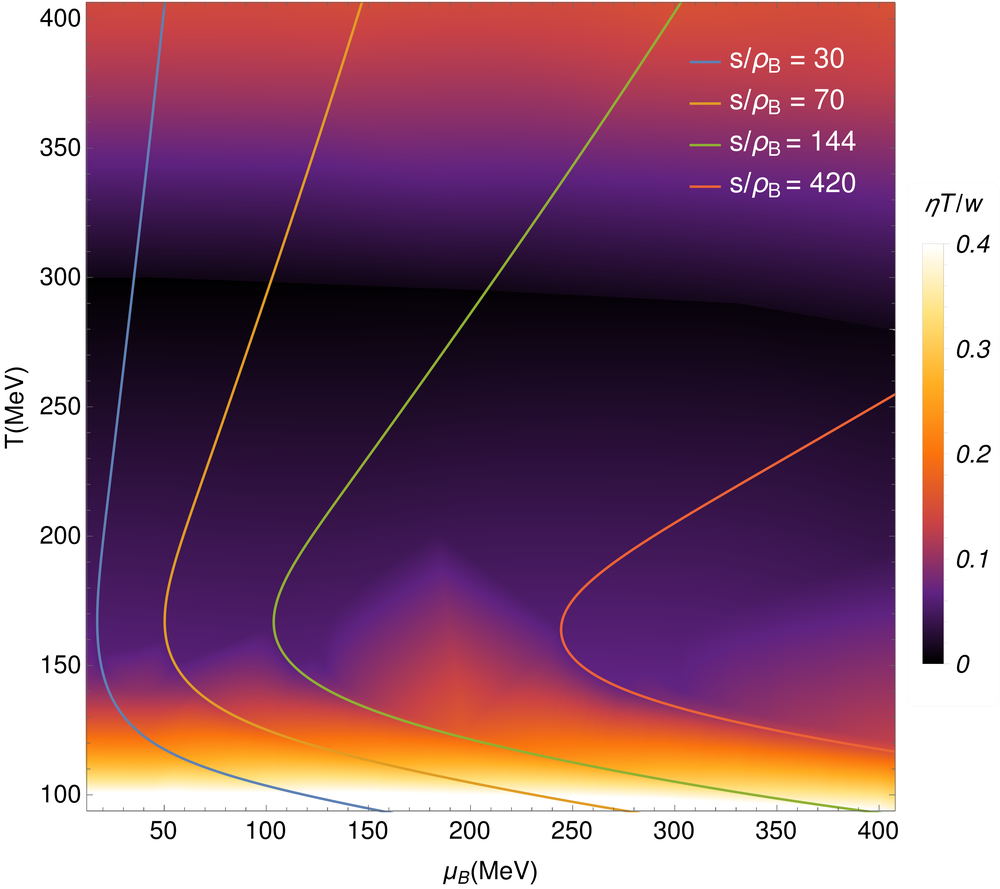}\\[2pt]
    
  \end{minipage}\hfill
  \begin{minipage}{0.49\columnwidth}
    \centering
    \includegraphics[width=\linewidth]{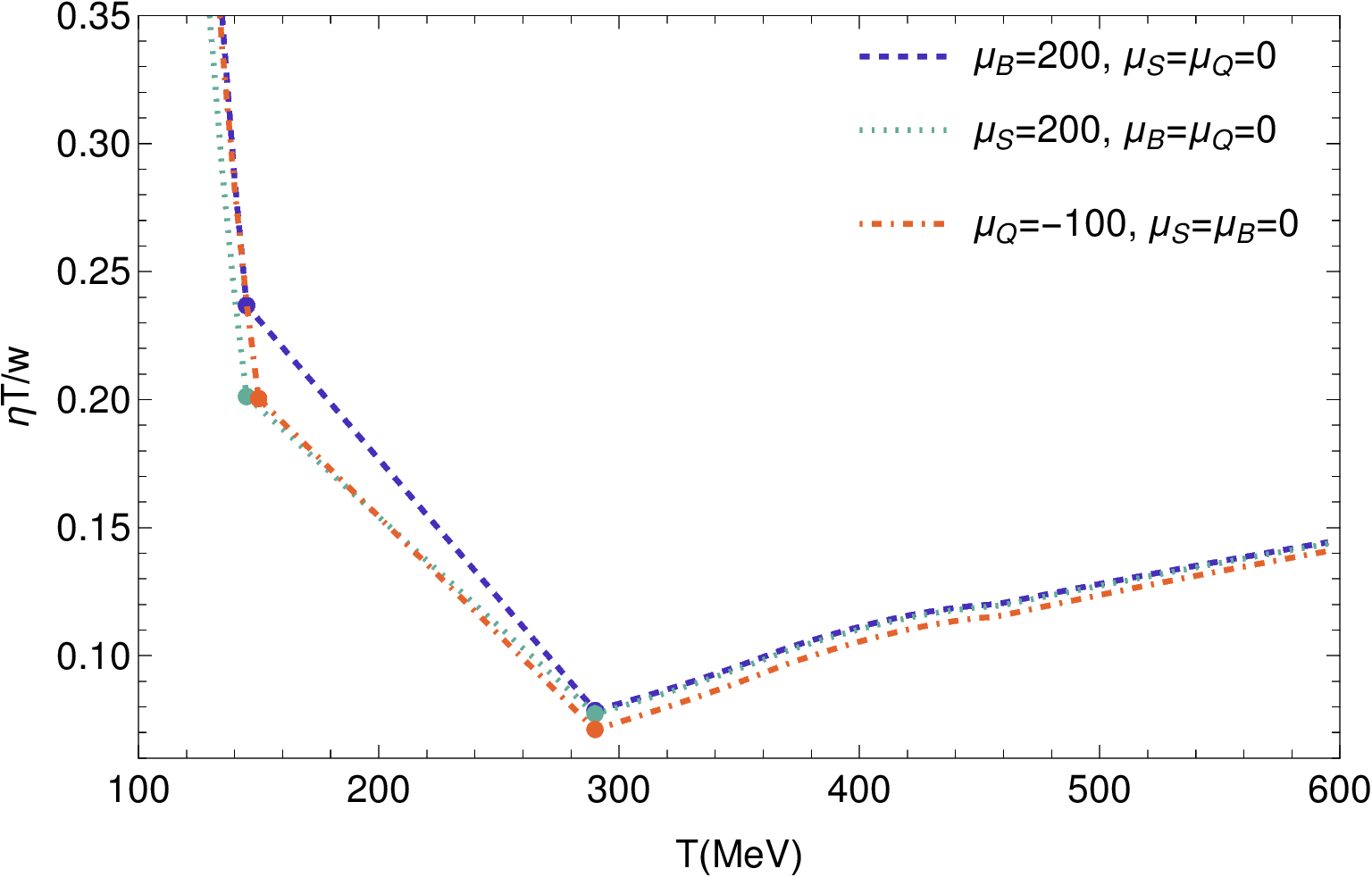}\\[2pt]
    
  \end{minipage}
  \caption{\textbf{(a):} $\eta T/w$ with $r=0.25$ fm for isentropes; colors indicate isentropes defined by $s/\rho_B$.\textbf{(b):} $\eta T/w$ vs.\ $T$ for $r=0.25$ fm; three chemical-potential scenarios, dots show interpolation region. Plots taken from \cite{Danhoni:2024ewq}}
  \label{fig:etaTw_side_by_side}
  
\end{figure}
Next, we combine these calculations using an interpolation function. 
Since shear viscosity for $T>300$ MeV was obtained using kinetic theory as a function of the BSQ chemical potentials and the strong coupling $1/(g^4 \log(g^{-1}))$, we first perform a rescaling, such that for $\mu=0$ our calculations match the NLO results from \cite{Ghiglieri:2018dib}. 
Once the rescaling is done, the interpolation function can be calculated by performing a matching between pQCD and HRG results. 
In this work, we choose to use a simple polynomial fit, and ensure that the transition points must always match such that $\eta T/w$ is continuous and $\eta T/w > 0$. Since the overall magnitude is significantly too high compared to what one expected for $\eta/s(T)$, in the following, we also implement an overall normalization constant $g_\mathrm{norm}$, and explore some interesting behaviors at finite $\tilde{\mu}$.
\begin{equation}
\left(\frac{\eta T}{w} \right)_\mathrm{tot}\left(T,\tilde{\mu}\right)
=
g_\mathrm{norm}
\begin{cases}
\left(\dfrac{\eta T}{w} \right)_\mathrm{HRG},
& T \leq T_\mathrm{sw}^\mathrm{HRG}, \\[6pt]
\left(\dfrac{\eta T}{w} \right)_\mathrm{intermediate},
& T_\mathrm{sw}^\mathrm{HRG} < T < T_\mathrm{sw}^\mathrm{pQCD}, \\[6pt]
g_\mathrm{GMT}\left(\dfrac{\eta T}{w} \right)_\mathrm{pQCD},
& T \geq T_\mathrm{sw}^\mathrm{pQCD}.
\end{cases}
\end{equation}

where $g_\mathrm{GMT}$ is the scaling factor to reproduce the pQCD results from \cite{Ghiglieri:2018dib} at $\tilde{\mu}=0$. In Fig.\ref{fig:etaTw_side_by_side}, we show the functional form obtained after following this procedure for different combinations of chemical potentials.

\vspace{-0.24cm}
\section{Next-to-leading order corrections for QCD}
\label{sec-5}
\begin{figure}[htbp]
  \centering
  \begin{minipage}{0.45\columnwidth}
    \centering
    \includegraphics[width=\linewidth]{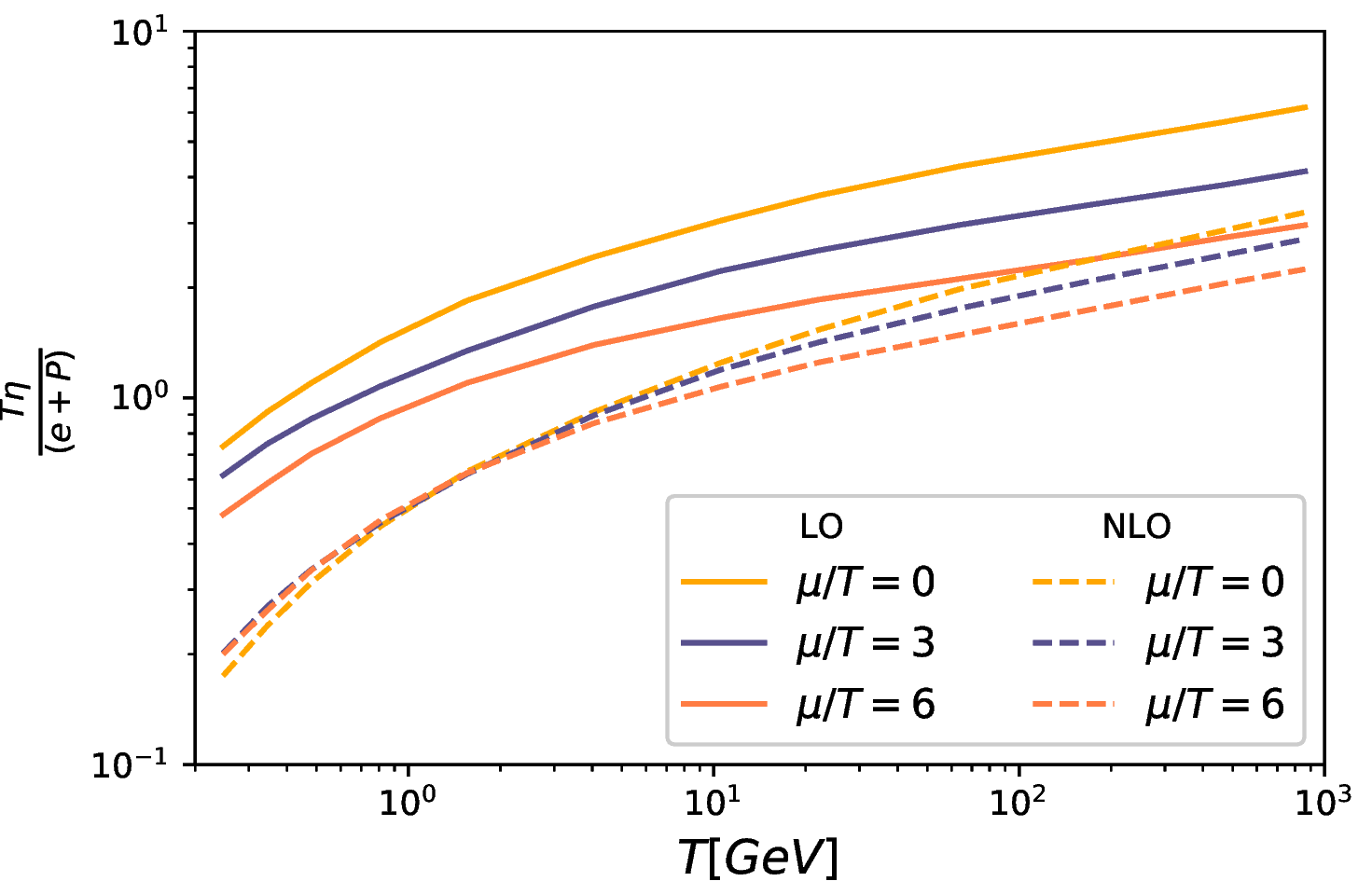}\\[2pt]
    \end{minipage}
  \begin{minipage}{0.45\columnwidth}
    \centering
    \includegraphics[width=\linewidth]{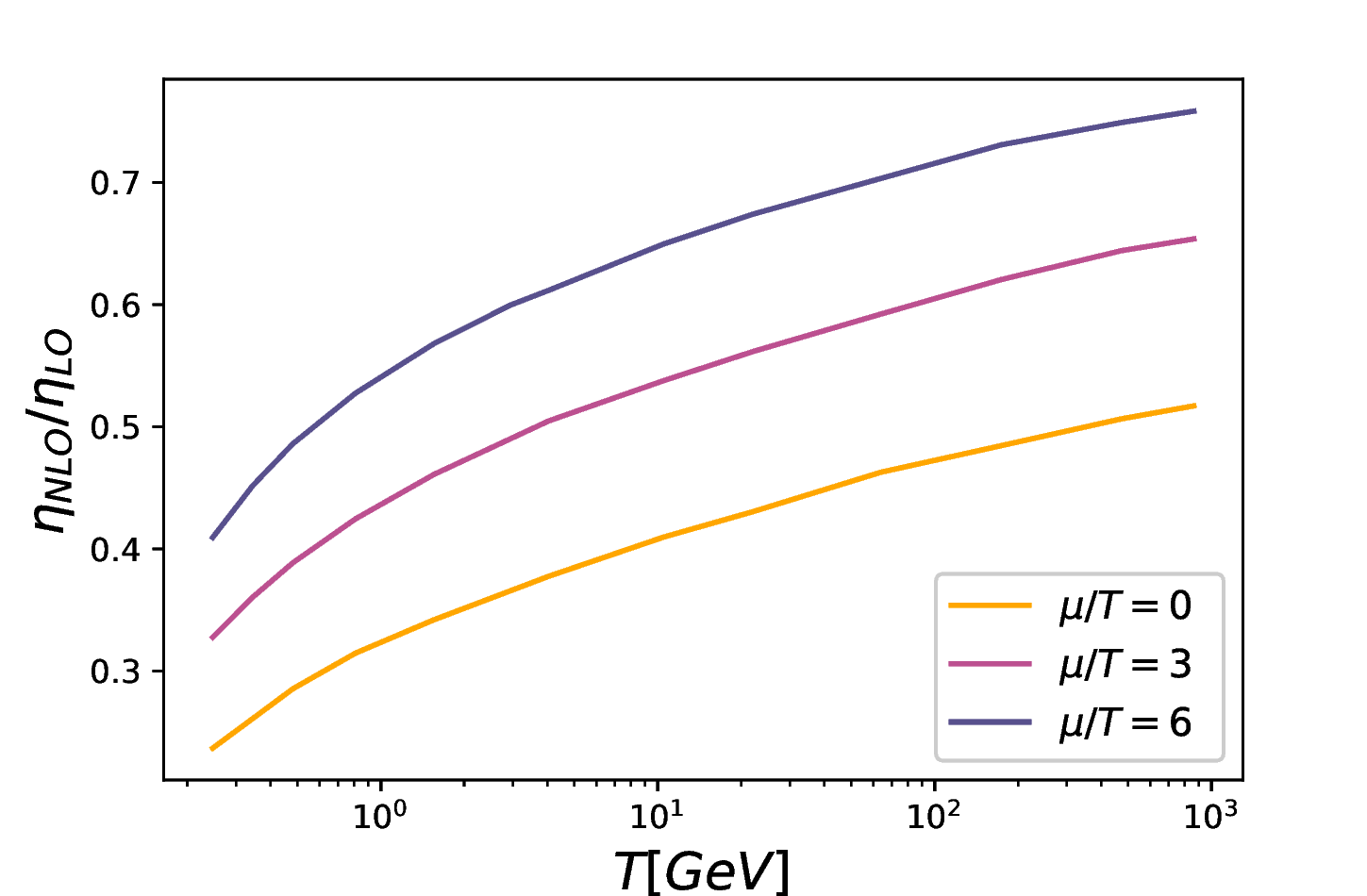}\\[2pt]  
  \end{minipage}
  \caption{\textbf{(a):} Kinematic shear viscosity vs.\ $T$ for $\mu=0,3,6$. Dashed: NLO, solid: LO.\textbf{(b):} Ratio $\eta_{\mathrm{NLO}}/\eta_{\mathrm{LO}}$ for three $\mu$ across $T$. $\muEQCD=2.7T$. Plots taken from \cite{Danhoni:2024ewq}.}
  \label{fig:visc_pair}
  \vspace{-0.7cm}
\end{figure}

\vspace{-0.2cm}
In this section, we present shear viscosity results for pQCD at finite $\mu$ at leading order (LO) and (almost) next-to-leading order (NLO), where the later includes effects from both LO and NLO; details are in Ref.~\cite{Danhoni:2024ewq}. In Fig.\ref{fig:visc_pair}, we show the kinematic viscosity, $\eta T/w$, on the left, and the ratio between LO and NLO results on the right. We note that $\mu$ in this case represents the quark chemical potential, which is taken to be the same for all quark species. 

We observe that NLO effects are rather important at all temperatures and chemical potentials considered in this work. However, the difference between LO and NLO is smaller for $\mu>0$ than at $\mu=0$, which is clearer at high temperatures, where the coupling is smaller and $\,\nf$ is larger. At $\mu=6T$, we can see a much better agreement even for relatively low temperatures than at $\mu/T=0$.
Additionally, we observe that the kinematic viscosity $\eta T/(e+P)$ is weakly dependent on $\mu$. The curve for large chemical potential falls below the $\mu=0$ result from \cite{Ghiglieri:2018dib}, except for $T=1$ GeV, where the $\mu/T$ dependence seems to vanish, and all curves agree.
Our results show that perturbation theory does not converge at experimentally achievable combinations of $(\mu, T)$, but there is a clear improvement at large $\mu$.

\vspace{-0.5cm}
\section{Conclusions}
\label{sec-6}
We combined pQCD results (with 3 conserved charges) and an excluded-volume HRG, together with a state-of-the-art list of resonances, to study the QCD shear viscosity at finite density. We have applied a phenomenological approach to produce curves of $\eta T/w(T,\mu_B,\mu_S,\mu_Q)$ across the QCD phase diagram, which can be used to feed relativistic viscous hydrodynamic codes simulating
collisions at energies covered by the RHIC Beam Energy Scan or BSQ fluctuations of conserved
charges at the LHC. The interplay of the three chemical potentials and the transition region leads to non-monotonic behavior in $\eta T/w$. Additionally, (almost) NLO results for $\eta T/w(\mu_B)$ show an improvement in the convergence of the perturbative series at large $\mu/T$, though NLO corrections remain quantitatively important.
\vspace{-0.5cm}
\section{Acknowledgments}
This research was
partly supported by the US-DOE Nuclear Science Grant No. DE-SC0023861 and by
the National Science Foundation (NSF) within the framework of the MUSES collaboration, under grant number
OAC-2103680.
\vspace{-0.5cm}
\bibliography{inspire,NOTinspire,another}

\end{document}